\documentclass[aps,prd,superscriptaddress,showpacs,preprintnumbers,10pt]{revtex4}
\usepackage[dvips]{graphicx,epsfig,color}
\usepackage{wrapfig,rotating}
\usepackage{amssymb,amsmath,array}

\usepackage{color}

\newcommand{\be}{\begin{equation}}
\newcommand{\ee}{\end{equation}}
\newcommand{\bq}{\begin{eqnarray}}
\newcommand{\eq}{\end{eqnarray}}

\newcommand{\D}{\mathrm{d}}

\def\Vec#1{\mathpalette{\VVec}{#1}}                  % size-respecting bold
\def\VVec#1#2{\mbox{\boldmath$#1#2$\unboldmath}}

\def\anti#1{\mathpalette{\@anti}{#1}#1}%              % better "anti" bar
\def\@anti#1#2{\sbox0{$#1#2$}%                        % auxiliary for above
  \makebox[0pt][l]{$#1\kern.30\ht0\overline{\kern-.35\ht0\phantom{#2}}$}}

\begin{document}

\title{Point-by-point extraction of parton distribution functions \\
from SIDIS single transverse--spin asymmetries}

\author{Anna~Martin}

\affiliation{Dipartimento di Fisica, Universit{\`a}
degli Studi di Trieste, 34127 Trieste, Italy}
\affiliation{INFN, Sezione di Trieste, 34127 Trieste, Italy}

\author{Franco~Bradamante} 
\affiliation{INFN, Sezione di Trieste, 34127 Trieste, Italy}

\author{Vincenzo~Barone}

\affiliation{Di.S.I.T., Universit{\`a} del Piemonte
Orientale ``A. Avogadro'', 15121 Alessandria, Italy; \\
INFN, Sezione di Torino,  10125 Torino, Italy}

\begin{abstract}
We show how some parton distribution functions related to 
the transverse spin of nucleons can be extracted 
point by point from combinations of proton and 
deuteron observables. In particular, we present 
a determination of the valence and sea Sivers functions 
from the single-spin asymmetries measured by COMPASS.

\end{abstract}

\pacs{13.88.+e, 13.60.-r, 13.66.Bc, 13.85.Ni}

\maketitle

% \section{Introduction}

The transverse--spin structure of the nucleon is presently one of the most relevant topics 
of hadronic physics (for reviews, 
see \cite{Barone:2002,bbm,Aidala:2013,EPJA}). On the experimental side, 
the semi-inclusive deep inelastic 
scattering (SIDIS) measurements have provided a wealth of data on single spin asymmetries, 
which shed light on the transversity distribution and on the leading-twist transverse-momentum dependent 
distribution functions (TMDs). 
In most phenomenological studies, these data are analyzed using specific functional forms 
for the transversity and the TMDs, with a certain number of free parameters 
determined by fits to the measured asymmetries. 
Alternatively, one can  adopt a simpler approach 
consisting in using simultaneously the proton and deuteron
asymmetries measured at the same $x$ and 
$Q^2$,  and performing a point-by-point extraction 
of the parton distribution functions directly from the data, 
with a very limited set of assumptions. 

\begin{figure*}[tb]
\centering
\includegraphics[width=0.45\textwidth]{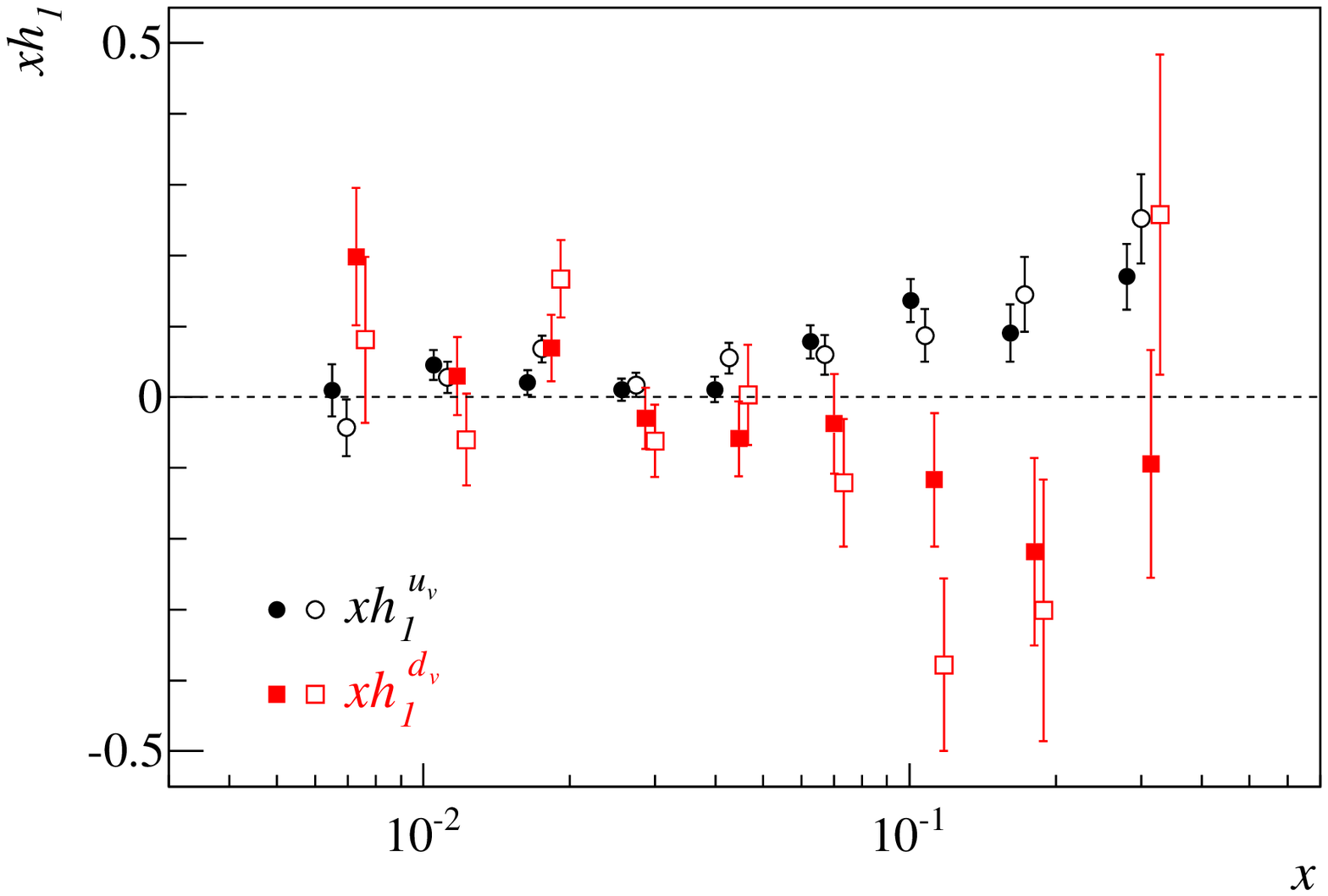}
\includegraphics[width=0.45\textwidth]{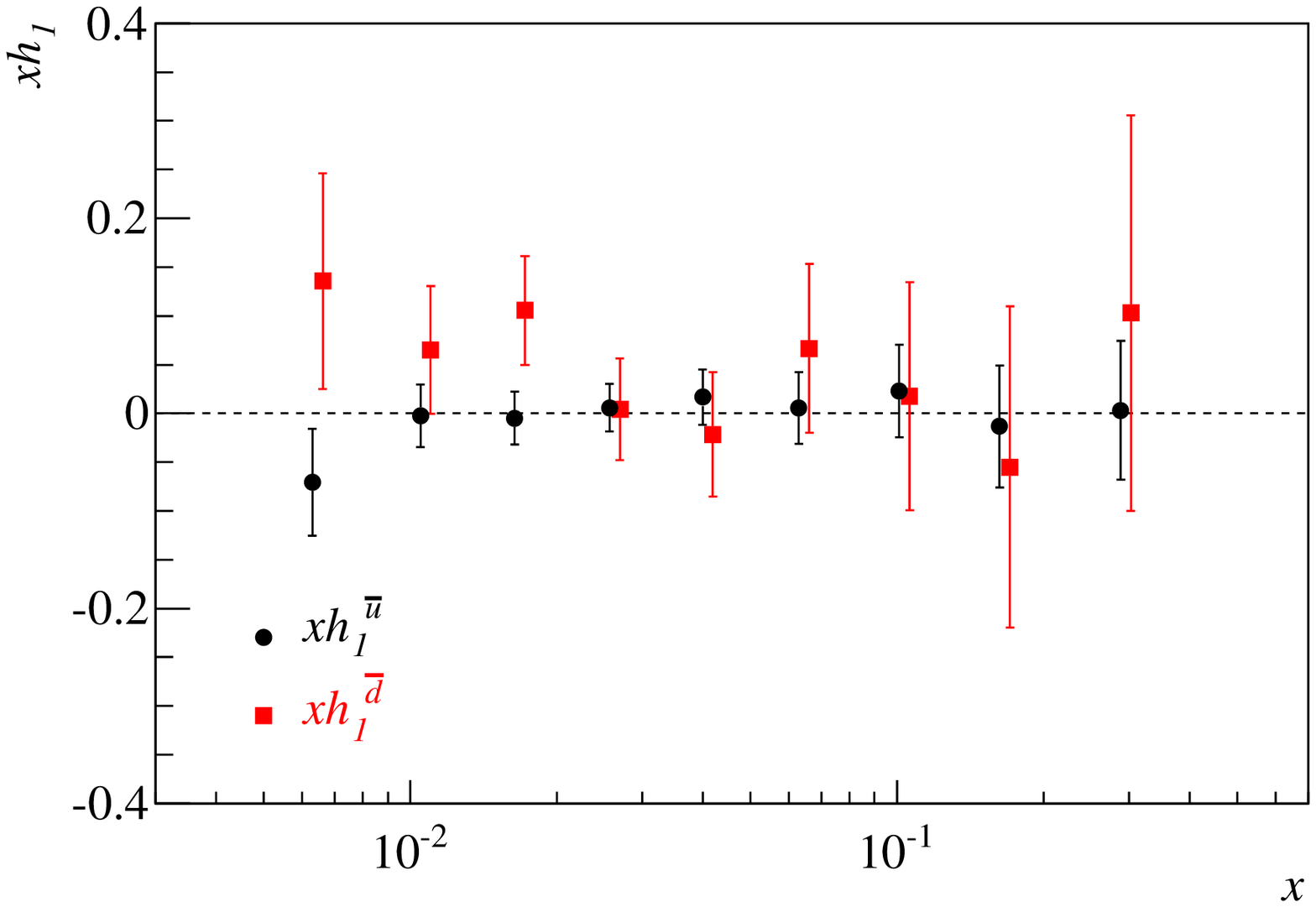}
\caption{
Left: the valence transversity distributions $xh^{u_v}_1$
and $xh^{d_v}_1$ from the COMPASS
dihadron (open points) 
and Collins asymmetries (solid points).
Right: the sea transversity distributions $xh^{\bar{u}}_1$
and $xh^{\bar{d}}_1$. Both plots are from Ref. \cite{Martin:2015}.}
\label{fig:h1}
\end{figure*}

In \cite{Martin:2015} we applied this method to extract 
the transversity distributions 
from the Collins and di-hadron asymmetries 
on proton and deuteron measured by the COMPASS Collaboration, 
using also the corresponding $e^+e^-$ asymmetries
from the Belle experiment. The results of our determination  
of the valence and sea transversity are shown in Fig.~\ref{fig:h1}. 
%

%%%%%%%%%%%%%%%%%%%%%%%%%%%%%%%%%%%%%%%%

The same method can be used to extract a very important TMD, the  
 Sivers function $f_{1T}^{\perp}$ \cite{Sivers:1990,Sivers:1991,Brodsky:2002,Collins:2002}, 
which encodes the correlation between the transverse momentum 
$\Vec{k}_T$
of quarks in a transversely polarized nucleon and
the spin of the parent nucleon. Here we briefly report on this extraction, 
presented in detail in \cite{Martin:2017yms}.

The asymmetry related to the Sivers function 
has been found to be different from zero for positive charge
hadrons produced on protons first by the HERMES experiment
\cite{Airapetian:2005,Airapetian:2009}
and a few years later,
at higher beam energy, by the COMPASS experiment
\cite{Alekseev:2009,Alekseev:2010rw,Adolph:2012,Adolph:2015}.
The first COMPASS measurements, performed using a deuteron target, showed no
clear signal
\cite{Alexakhin:2005,Ageev:2006da,Alekseev:2009}.
Measurements on pion production on a transversely polarized $^3He$ target 
and 6 GeV electron beam have been performed 
more recently by the Hall A Collaboration at JLab~\cite{Qian:2011}.

The Sivers asymmetry is proportional to a convolution over 
transverse momenta of the Sivers function $f_{1T}^{\perp}$ and of the unpolarized 
fragmentation function $D_1$. It can be factorized 
using a Gaussian Ansatz \cite{Boer:1998,Efremov:2003,Efremov:2005} and becomes
\be
A_h(x,z, Q^2) 
= G \,   \frac{\sum_{q,\bar{q}} e_q^2 x f_{1T}^{\perp (1) q}(x, Q^2) z D_{1q}(z, Q^2)}
{\sum_{q,\bar{q}} e_q^2 x f_1^q(x, Q^2)  D_{1q}(z, Q^2)}
\label{sivers22}
\ee
where 
\be
f_{1T}^{\perp (1)}(x, Q^2) \equiv \int \D^2  \Vec k_T 
\, \frac{k_T^2}{2 M^2} \,  
\, f_{1T}^{\perp}(x, k_T^2, Q^2)  
\label{k_T_moment}
\ee
is the first $k_T^2$ moment of the Sivers function. 
The $G$ factor, resulting from the  integration over transverse
momenta,  is given by
\be
G  =  \frac{\sqrt{\pi} M}{\sqrt{\langle p_T^2 \rangle + 
z^2 \langle k_T^2 \rangle_S}},  
\ee
where $\langle p_T^2 \rangle$ and $\langle k_T^2 \rangle_S$ 
are the widths of the transverse-momentum parts of the 
 fragmentation function and of the Sivers function 
respectively.
A good approximation 
is to set $G \simeq \pi M / 2 \langle P_{h \perp} \rangle$,
where $\langle P_{h \perp} \rangle$ is the mean value
of the  final hadron transverse momentum, and take it as a constant, 
since the measured $z$ dependence  of $\langle P_{h \perp} \rangle$ 
is smooth in the range of interest. 
This approximation, which should give systematic corrections
well within the overall uncertainties, allow us 
to the Sivers asymmetry as a function of $x$ as
\be
A_{h}(x, Q^2) = G \,  
\frac{\sum_{q,\bar{q}} e_q^2 x f_{1T}^{\perp (1) q}(x, Q^2)  \widetilde{D}_{1q}^{(1)}(Q^2)}
{\sum_{q,\bar{q}} e_q^2 x f_1^q(x, Q^2)  \widetilde{D}_{1q} (Q^2)}. 
\label{eq:sivers}
\ee
where
\bq
\widetilde{D}_{1}(Q^2) = 
\int \D z \, D_{1} (z, Q^2)\,, \; \; \;
 \widetilde{D}_{1}^{(1)}(Q^2) =
\int \D z \, z D_{1}(z, Q^2)\, .
\eq

The fragmentation functions and the unpolarized distribution functions 
appearing in eq.~(\ref{eq:sivers})  
can be obtained from standard parametrizations, so  
we can extract the transverse moments of the Sivers 
function $f_{1T}^{\perp (1)}$ by properly combining 
the asymmetries on proton and deuteron, for charged
pions and kaons.

In the pion case we use the
favored and unfavored fragmentation functions defined as
\bq
D_{1, {\rm fav}}^{\pi} &\equiv& D_{1u}^{\pi^+} = D_{1d}^{\pi^-} = D_{1\bar{u}}^{\pi^-} 
= D_{1\bar{d}}^{\pi^+} \, , \nonumber \\
D_{1, {\rm unf}}^{\pi} &\equiv& D_{1u}^{\pi^-} = D_{1d}^{\pi^+} = D_{1\bar{u}}^{\pi^+} 
= D_{1\bar{d}}^{\pi^-} \, ,
\label{eq:favdis2}
\eq
and for the strange quark we assume
\be
D_{1s}^{\pi^\pm} = D_{1 \bar s}^{\pi^\pm} = N \, D_{1, {\rm unf}}^{\pi}  \, , 
\ee
with the constant factor $N \simeq 0.8$ evaluated in \cite{deflorian}.
The asymmetries can then be expressed in terms of the ratios of fragmentation 
functions
\bq
\beta_{\pi} (Q^2) = \frac{\widetilde{D}_{1, {\rm unf}}^{\pi}(Q^2)}{\widetilde{D}_{1, {\rm fav}}^{\pi}
 (Q^2)} , \; \; \;
\beta_{\pi}^{(1)}(Q^2) = \frac{\widetilde{D}_{1, {\rm unf}}^{\pi (1)}(Q^2)}
{\widetilde{D}_{1, {\rm fav}}^{\pi (1)}(Q^2)} \, ,  \; \; \;
\rho_{\pi} (Q^2) = \frac{\widetilde{D}_{1, {\rm fav}}^{\pi (1)}(Q^2)}
{\widetilde{D}_{1, {\rm fav}}^{\pi} (Q^2)}\,,  
\label{mom_ratio}
\eq
and the valence Sivers distributions turn out to be given by 
\bq
x f_{1T}^{\perp (1) u_v} &=& \frac{1}{5 G \rho_{\pi} (1 - \beta_{\pi}^{(1)})} 
\left [ ( x f_p^{\pi^+} A_p^{\pi^+} - x f_p^{\pi^-} A_p^{\pi^-})
+\frac{1}{3}  ( x f_d^{\pi^+} A_d^{\pi^+} - x f_d^{\pi^-} A_d^{\pi^-})  \right ] \,, 
\nonumber \\
x f_{1T}^{\perp (1) d_v} &=&  \frac{1}{5 G \rho_{\pi} (1 - \beta_{\pi}^{(1)})} 
\left [ \frac{4}{3}  ( x f_d^{\pi^+} A_d^{\pi^+} - x f_d^{\pi^-} A_d^{\pi^-})
-(x f_p^{\pi^+} A_p^{\pi^+} - x f_p^{\pi^-} A_p^{\pi^-}) \right ]\,,
\label{dval_siv}
\eq
where $f_{p,d}^{\pi^{\pm}}$ are linear combinations of the unpolarized distribution 
functions (for their explicit expressions see 
\cite{Martin:2017yms}). From the measured asymmetries one can also obtain directly 
the difference of the sea distributions  
$ x f_{1T}^{\perp (1) \bar u} - x f_{1T}^{\perp (1) \bar d}$:  
 \bq
  x f_{1T}^{\perp (1) \bar u} -  x f_{1T}^{\perp (1) \bar d} &=& \frac{1}{15 G \rho_{\pi} 
 \left (1 - \beta_{\pi}^{(1) 2} \right )} 
\,  \left [ 2 (1 - 4 \beta_{\pi}^{(1)}) x f_p^{\pi^+} A_p^{\pi^+} 
 + 2 (4 - \beta_{\pi}^{(1)}) x f_p^{\pi^-} A_p^{\pi^-} 
 \right. \nonumber \\
 & & - \left. (1 - 4 \beta_{\pi}^{(1)}) x f_d^{\pi^+} A_d^{\pi^+} 
 - (4 - \beta_{\pi}^{(1)}) x f_d^{\pi^-} A_d^{\pi^-})  \right ] \,.  
 \label{ubar-dbar}
 \eq

In the case of charged kaons, following the same procedure,
we introduce
\begin{eqnarray}
D_{1, {\rm fav}}^K &\equiv& D_{1u}^{K^+}  = D_{1\bar{u}}^{K^-} \, , \; \;
D_{1, {\rm fav}}'^K \equiv D_{1\bar{s}}^{K^+} = D_{1 s}^{K^-} ,
\nonumber \\
D_{1, {\rm unf}}^K &\equiv& D_{1d}^{K^\pm} = D_{1\bar{d}}^{K^\pm} = D_{1\bar{u}}^{K^+} 
= D_{1 u}^{K^-} = D_{1s}^{K^+} = D_{1 \bar s}^{K^-} , 
\end{eqnarray}
the ratios $\beta_{K}$, $\beta_{K}^{(1)}$ and $\rho_{K}$ defined as in
eq.~(\ref{mom_ratio}), but for kaons, and
\be
\gamma_{K} (Q^2) = \frac{\widetilde{D}_{1, {\rm fav}}'^{K}(Q^2)}{\widetilde{D}_{1, {\rm fav}}^{K}
 (Q^2)}, 
\; \;
\gamma_{K}^{(1)} (Q^2) = \frac{\widetilde{D}_{1, {\rm fav}}'^{K (1)}(Q^2)}{\widetilde{D}_{1, 
{\rm fav}}^{K (1)}
 (Q^2)} \, .
\label{beta_gamma_K(1)}
\ee
\begin{figure*}[t]
\centering
\includegraphics[width=0.45\textwidth]{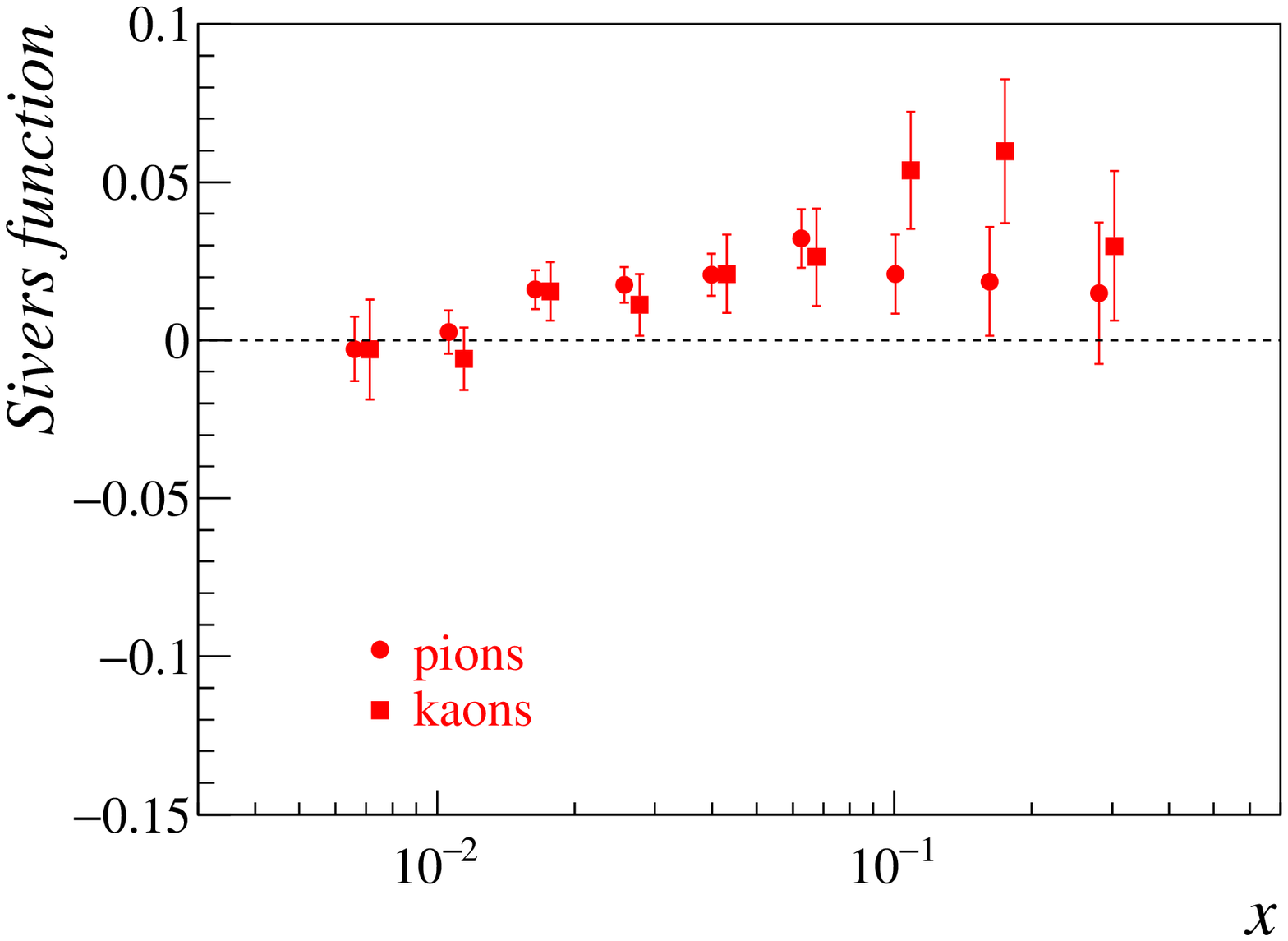}
\includegraphics[width=0.45\textwidth]{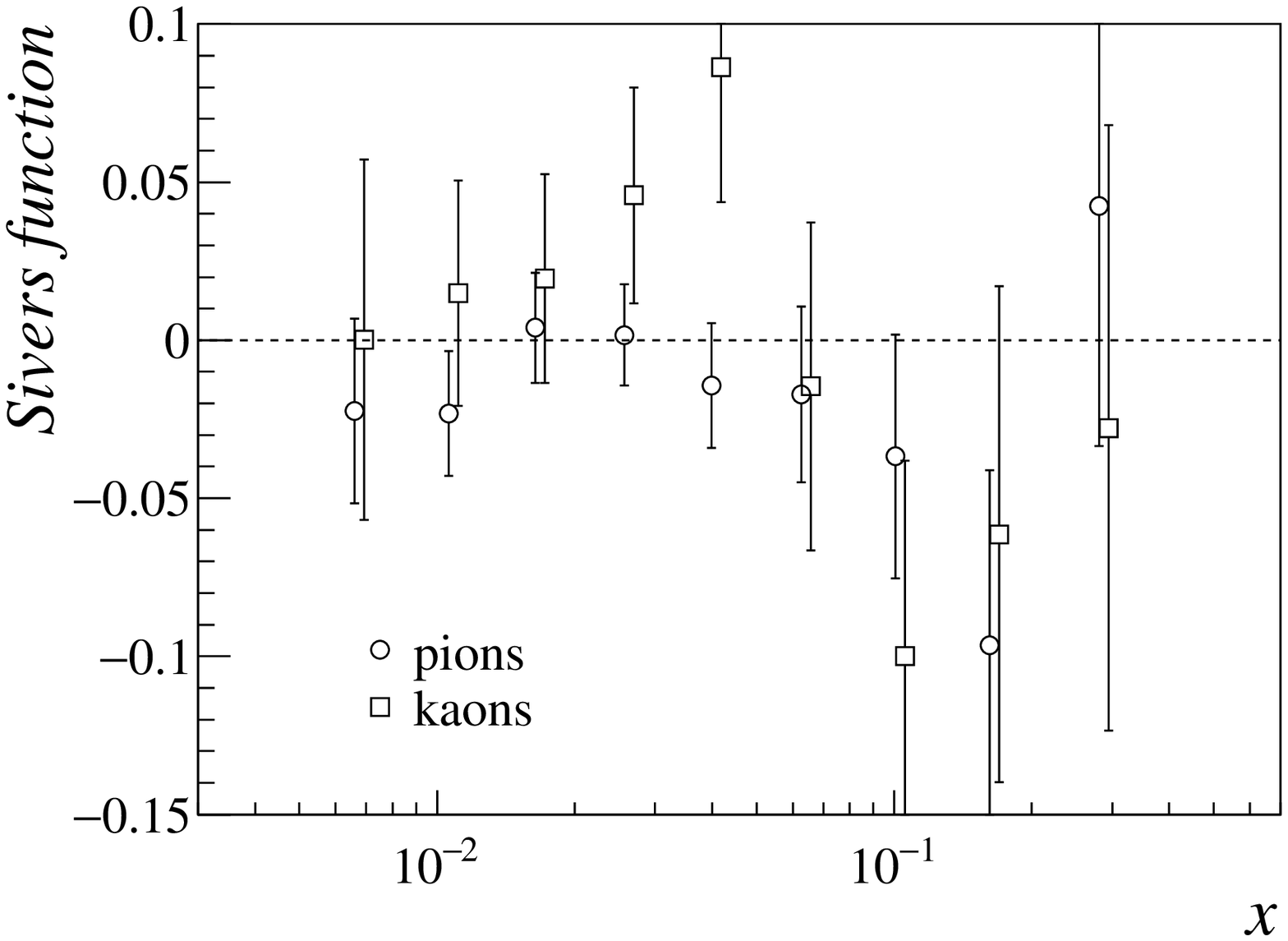}
\caption{Comparison of the first $k_T^2$ moments of the Sivers valence distributions, 
$x f_{1T}^{\perp (1) u_v}$ (left) and $x f_{1T}^{\perp (1) d_v}$ (right),  
obtained from pion (dots) and kaon (squares) data.  
}
\label{fig:val_ud}
\end{figure*}
In this case, assuming
that the difference of strange sea distributions
$x f_{1T}^{\perp (1) s} - x f_{1T}^{\perp (1) \bar s}$ 
is negligible, we obtain
\bq
x f_{1T}^{\perp (1) u_v}  &= &
\frac{1}{4 G \rho_{K} (1 - \beta_{K}^{(1)})} 
\left [ ( x f_p^{K^+} A_p^{K^+} - x f_p^{K^-} A_p^{K^-}) \right ]
% \label{uval_siv_K}
\nonumber \\
x f_{1T}^{\perp (1) d_v} &=&  \frac{1}{4 G \rho_{K} (1 - \beta_{K}^{(1)})} 
\left [  ( x f_d^{K^+} A_d^{K^+} - x f_d^{K^-} A_d^{K^-}) 
- (x f_p^{K^+} A_p^{K^+} - x f_p^{K^-} A_p^{K^-}) \right ]\,, 
\label{dval_siv_K}
\eq
where the quantities $f_{p,d}^{K^{\pm}}$ are linear combinations of the 
unpolarized distribution functions.  

To extract the Sivers functions from eqs.~(\ref{dval_siv}), 
(\ref{ubar-dbar}) and (\ref{dval_siv_K}) we used the COMPASS measurements of 
the Sivers asymmetries in SIDIS of 160 GeV muons
on proton \cite{Adolph:2015} and deuteron targets \cite{Alekseev:2009}
for charged pions and kaons.
The $x$ binning is the same for all series of data. 
Concerning the momentum transfer $Q^2$, 
it ranges from 1.2 GeV$^2$ for the lowest $x$ point to 20 GeV$^2$ 
for the highest $x$ point. 
We have used 
the unpolarized distribution functions
from the CTEQ5D global fit \cite{cteq}, and the unpolarized fragmentation functions from 
the DSS parametrization \cite{deflorian}. 
Finally, the quantity $G=\pi M / 2 \langle P_{h \perp} \rangle$
has been calculated using the measured $\langle P_{h \perp} \rangle$, 
which is $\sim 3$ for pions and $\sim 2.5$ for 
kaons with a slight $x$ dependence.

Figure~\ref{fig:val_ud} shows the extracted values
of the Sivers distribution $x f_{1T}^{\perp (1) u_v}$ 
(left)
and $x f_{1T}^{\perp (1) d_v}$ (right),
as obtained from pion and kaon data.
The error bars indicate the statistical uncertainties only.
The $u_v$ distribution is clearly positive and different from zero
over most of the covered $x$ range.
The statistical errors for $x f_{1T}^{\perp (1) d_v}$ are much larger
because of the unbalanced proton--deuteron statistics in the
COMPASS data.
Still, the $d_v$ distribution appears to be negative in the valence
region and the values are compatible with 
$x f_{1T}^{\perp (1) d_v} \simeq - x f_{1T}^{\perp (1) u_v}$.
The agreement between the independent results obtained from pion and 
kaon data is
quite good, as expected. Also, our results agree rather well with
previous extractions (for instance, with the fits 
of \cite{Anselmino:2009,Anselmino:2012}). 

\begin{figure}[t]
\centering
\includegraphics[width=0.45\textwidth]{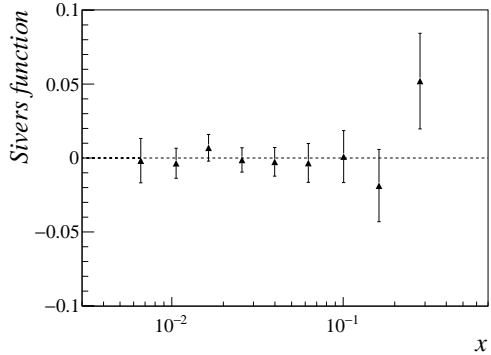}
\caption{The isotriplet Sivers sea $x f_{1T}^{\perp (1) \bar u} - x f_{1T}^{\perp (1) \bar d}$ 
extracted from pion asymmetry data.}
\label{fig:ubmdb}
\end{figure}

The sea difference 
$x f_{1T}^{\perp (1) \bar u} - x f_{1T}^{\perp (1) \bar d}$
obtained  from the pion asymmetries is shown in Fig.~\ref{fig:ubmdb}: 
as one can see, it is 
compatible with zero, with small statistical
uncertainties. 

To summarize, the first $k_T^2$ moments of the Sivers distributions have
been extracted directly from the Sivers asymmetries for
charged pions and kaons measured by COMPASS using proton and
deuteron targets in the same kinematical region and 
in the same $x$ bins.
The main advantage of this
point-by-point determination 
is that no specific parametrization
of $f_{1T}^{\perp (1)}$ is required.
Our results  clearly show a non-vanishing and positive $u_v$ Sivers function 
different from zero, and an
isotriplet Sivers sea $x f_{1T}^{\perp (1) \bar u} - x f_{1T}^{\perp (1) \bar d}$
compatible with zero. 
As for the $d_v$ Sivers function, it has opposite sign
with respect to the $u_v$ distribution, but to improve its knowledge 
more precise deuteron data are clearly needed.


\begin{thebibliography}{10}

\bibitem{Barone:2002}
V.~Barone, A.~Drago, and P.~G. Ratcliffe,
 Phys. Rep. {\bf 359}, 1 (2002).

\bibitem{bbm}
V.~Barone, F.~Bradamante and A.~Martin, 
Progr. Part. Nucl. Phys. {\bf 65}, 267 (2010). 

\bibitem{Aidala:2013} 
C.A.~Aidala, S.D.~Bass, D.~Hasch, and G.K.~Mallot, Rev. Mod. Phys. {\bf 85}, 655 (2013). 

\bibitem{EPJA}
{\it The 3-D Structure of the Nucleon}, special issue of Eur. Phys. J. A {\bf 52}, no. 6, 2016.  


\bibitem{Martin:2015}
A.~Martin, F.~Bradamante, and V.~Barone, Phys. Rev. D {\bf 91}, 014034 (2015). 



\bibitem{Sivers:1990}
D.~Sivers, Phys. Rev. D {\bf 41}, 83 (1990).

\bibitem{Sivers:1991}
D.~Sivers, Phys. Rev. D {\bf 43}, 261 (1991).

\bibitem{Brodsky:2002} 
S.J.~Brodsky, D.S.~Hwang, and I.~Schmidt, Phys. Lett. B {\bf 530}, 99 (2002).


\bibitem{Collins:2002} 
J.C.~Collins, Phys. Lett. B {\bf 536}, 43 (2002)


%\cite{Martin:2017yms}
\bibitem{Martin:2017yms}
  A.~Martin, F.~Bradamante and V.~Barone,
  %``A direct extraction of the Sivers distributions from spin asymmetries in pion and kaon leptoproduction,''
  arXiv:1701.08283 [hep-ph].



\bibitem{Airapetian:2005}
A.~Airapetian {\it et al.} (HERMES Collaboration), Phys. Rev. Lett. {\bf 94}, 
012002 (2005). 
 
\bibitem{Airapetian:2009}  
A.~Airapetian {\it et al.} (HERMES Collaboration), 
%Observation of the naive T-odd Sivers effect in deep-inelastic scattering
Phys. Rev. Lett. {\bf 103}, 152002 (2009). 

\bibitem{Alekseev:2009}
M.~Alekseev {\it et al.} (COMPASS Collaboration), 
%Collins and Sivers asymmetries for pions and kaons 
% in muon-deuteron DIS
Phys. Lett. B {\bf 673}, 127 (2009). 

\bibitem{Alekseev:2010rw} 
  M.~G.~Alekseev {\it et al.}  (COMPASS Collaboration),
  %``Measurement of the Collins and Sivers asymmetries on transversely polarised protons,''
  Phys.\ Lett.\ B {\bf 692}, 240 (2010).
  %  [arXiv:1005.5609 [hep-ex]].
  %%CITATION = ARXIV:1005.5609;%%

\bibitem{Adolph:2012}
C.~Adolph {\it et al.} (COMPASS Collaboration), 
% Experimental investigation of transverse spin asymmetries in mu-p SIDIS processes:
% Sivers asymmetries
Phys. Lett. B {\bf 717}, 107 (2012). 

\bibitem{Adolph:2015}
C.~Adolph {\it et al.} (COMPASS Collaboration), 
% Collins and Sivers asymmetries in muonproduction of pions and kaons 
% off transversely polarised protons
Phys. Lett. B {\bf 744}, 250 (2015). 



\bibitem{Alexakhin:2005} 
V.Yu.~Alexakhin {\it et al.} (COMPASS Collaboration), Phys. Rev. Lett. {\bf 94}, 202002 (2005). 

\bibitem{Ageev:2006da} 
  E.~S.~Ageev {\it et al.}  (COMPASS Collaboration),
  %``A New measurement of the Collins and Sivers asymmetries on a transversely polarised deuteron target,''
  Nucl.\ Phys.\ B {\bf 765}, 31 (2007).
%  [hep-ex/0610068].
  %%CITATION = HEP-EX/0610068;%%

\bibitem{Qian:2011} 
X.~Qian {\it et al.} (JLab Hall A Collaboration), 
Phys. Rev. Lett. {\bf 107}, 072003 (2011). 

\bibitem{Boer:1998}
D.~Boer and P.J.~Mulders, Phys. Rev. D {\bf 57}, 5780 (1998). 


\bibitem{Efremov:2003}
A.V.~Efremov, K.~Goeke, and P.~Schweitzer, 
Phys. Lett. B {\bf 568}, 63 (2003). 

\bibitem{Efremov:2005} 
A.V.~Efremov, K.~Goeke, S.~Menzel, A.~Metz, and P.~Schweitzer, 
Phys. Lett. B {\bf 612}, 233 (2005). 

%\bibitem{Collins:2005}
%J.C.~Collins, A.V.~Efremov, K.~Goeke, S.~Menzel, A.~Metz, and P.~Schweitzer, Phys. Rev. D {\bf 73}, 
%014021 (2006). 

%\bibitem{Vogelsang:2005}
%W.~Vogelsang, and F.~Yuan, Phys. Rev. D {\bf 72}, 054028 (2005).

%\bibitem{Anselmino:2005a}
%M.~Anselmino, M.~Boglione, U.~D'Alesio, A.~Kotzinian, F.~Murgia, and A.~Prokudin, 
%Phys. Rev. D {\bf 71}, 074006 (2005). 

%\bibitem{Anselmino:2005b}
%M.~Anselmino, M.~Boglione, U.~D'Alesio, A.~Kotzinian, F.~Murgia, and A.~Prokudin, 
%Phys. Rev. D {\bf 72}, 094007 (2005). 

%\bibitem{Anselmino:2006}
%M.~Anselmino {\it et al.}, in {\it Transversity 2005}, V.~Barone and P.G.~Ratcliffe, eds., 
%World Scientific, Singapore (2006), p.~236; arXiv:hep-ph/0511017. 

%\bibitem{Anselmino:2009}
%M.~Anselmino, M.~Boglione, U.~D'Alesio, A.~Kotzinian, S.~Melis, F.~Murgia, A.~Prokudin, and 
%C.~T{\"u}rk, Eur. Phys. J. A {\bf 39}, 89 (2009).  

%\bibitem{Anselmino:2012}
%M.~Anselmino, M.~Boglione, and S.~Melis, Phys. Rev. D {\bf 86}, 014028 (2012). 

%\bibitem{Sun:2013} 
%P.~Sun and F.~Yuan, Phys. Rev. D {\bf 88}, 034016 (2013). 

%\bibitem{Echevarria:2014}
%M.G.~Echevarria, A.~Idilbi, Z.-B.~Kang, and I.~Vitev, Phys. Rev. D {\bf 89}, 074013 (2014).  

\bibitem{deflorian}
D. de Florian, R.~Sassot and M.~Stratmann, Phys. Rev. D {\bf 75}, 114010 (2007). 



\bibitem{cteq} 
H.L.~Lai {\it et al.} (CTEQ Collaboration), Eur. Phys. J. {\bf C12}, 375 (2000). 


\bibitem{Anselmino:2009}
M.~Anselmino, M.~Boglione, U.~D'Alesio, A.~Kotzinian, S.~Melis, F.~Murgia, A.~Prokudin, and 
C.~T{\"u}rk, Eur. Phys. J. A {\bf 39}, 89 (2009).  

\bibitem{Anselmino:2012}
M.~Anselmino, M.~Boglione, and S.~Melis, Phys. Rev. D {\bf 86}, 014028 (2012). 




\end{thebibliography}
\end{document}